\documentclass[12pt]{article} 
\usepackage{graphicx,subfigure,latexsym}

\newcommand{\be}{\begin{equation}}
\newcommand{\ee}{\end{equation}}
\newcommand{\bear}{\begin{eqnarray}}
\newcommand{\eear}{\end{eqnarray}}
\newcommand{\ba}{\begin{array}}
\newcommand{\ea}{\end{array}}

\begin{document}

\begin{titlepage}
\vfill

\begin{flushright}
{\normalsize RBRC-1004}\\
\end{flushright}

\vfill
\begin{center}
{\Large\bf Interplay of Reggeon and photon in pA collisions}

\vskip 0.3in
Gokce Basar$^{1}$\footnote{e-mail: {\tt basar@uic.edu }}, 
Dmitri E. Kharzeev$^{2,3,4}$\footnote{e-mail:
{\tt dmitri.kharzeev@stonybrook.edu}},
Ho-Ung Yee$^{1,4}$\footnote{e-mail:
{\tt hyee@uic.edu}},
and Ismail Zahed$^{2}$\footnote{e-mail: {\tt ismail.zahed@stonybrook.edu }}
\vskip 0.15in

{\it $^{1}$ Department of Physics, University of Illinois, Chicago, Illinois 60607}\\[0.15in]
 {\it $^{2}$Department of Physics and Astronomy, Stony Brook University,} \\
{\it Stony Brook, New York 11794-3800 }\\[0.15in]
{\it $^{3}$Department of Physics, Brookhaven National Laboratory,} \\
{\it Upton, New York 11973-5000 }\\[0.15in]
{\it $^{4}$ RIKEN-BNL Research Center, Brookhaven National Laboratory,}\\
{\it Upton, New York
11973-5000}\\[0.15in]


\end{center}

\vfill

\begin{abstract}
We discuss the effects of the electromagnetic interaction in high-energy proton collisions with nuclei of large Z
at strong coupling $\lambda=g^2N_c$. Using the holographic dual limit of large $N_c>\lambda\gg 1$, we describe the Reggeon exchange as a twisted surface and show that it gets essentially modified by the electromagnetic interaction. 
\end{abstract}

\vfill

\end{titlepage}
\setcounter{footnote}{0}

\baselineskip 18pt \pagebreak
\renewcommand{\thepage}{\arabic{page}}
\pagebreak

\section{Introduction}

High energy diffractive hadron interactions with a large rapidity gap $\chi={\rm ln}(s/s_0)$ are dominated by
the Reggeon and Pomeron exchanges. In pp scattering the Reggeon exchange is dominant below $\sqrt{s}=5$ GeV, while
the Pomeron exchange dominates at higher $\sqrt{s}$ \cite{Donnachie:1992ny}. At weak coupling, the Pomeron and Reggeon
exchanges are described by rapidity ordered BFKL ladder diagrams~\cite{BFKL}. The effects of the electromagnetic interaction are usually assumed to be weak.

At strong t'Hooft coupling $\lambda=g^2N_c$, the Pomeron and Reggeon exchanges have been addressed
in the context of holography with and without supersymmetry by a number of authors~\cite{Rho:1999jm,Janik:2000aj,Janik:2001sc,Polchinski:2001tt,BLACK,Brower:2006ea,iancu,Alday:2007hr,KOV,Barnes:2009ag,Nishio:2011xz,Giordano:2011sn,Giordano:2011ua,Basar:2012jb,Stoffers:2012zw,ALEX-DIFF,Watanabe:2012uc,ALEX-ENTROPY,shuryak}. Since QCD is not supersymmetric, the Reggeon and Pomeron are best described by surface exchanges with non-supersymmetric holographic metric, whereby the massless spin-2 graviton transmutes to a massive spin-2 glueball and decouple in the
pertinent kinematics~\cite{Janik:2000aj,Janik:2001sc,Basar:2012jb,Stoffers:2012zw,ALEX-DIFF,ALEX-ENTROPY,shuryak}. The surface exchanges are noteworthy as they  encode a stringy Schwinger mechanism~\cite{Basar:2012jb}, which is also present when the
surface extrinsic curvature is included~\cite{Qian:2014jna}. They play an important role in the initial conditions for
both saturation~\cite{Qian:2014rda} and prompt thermalization~\cite{Shuryak:2013sra,Qian:2015boa}.

Our analysis below will revise and extend the picture of Reggeon exchange initially discussed~\cite{Janik:2001sc}, with a particular
focus on the role of electromagnetic corrections in pA collisions with a large Z nucleus. The dedicated pA experiments
at the LHC and RHIC colliders using heavy nuclei may access the physics that we will detail below. In section 2, we review the
analysis of the Reggeon amplitude at strong coupling using a semi-classical surface analysis. In section 3, we show how the effects of photon exchange modify the semi-classical analysis. 
Our conclusions follow in section 4.

\section{Reggeon amplitudes in strong coupling}

In this section, we give a brief review of the framework for computing Reggeon amplitudes in strong coupling, following the work of Ref.\cite{Janik:2001sc}.
The motivation of the set-up comes from the AdS/CFT correspondence or holographic QCD, where a strongly coupled, large $N_c$ gauge theory may be
described by a dual string theory in a holographic 5 dimensional space-time. The extra holographic dimension represents an energy scale of the problem, so that
it can be thought of as a geometric realization of renormalization group flow between different energy scales. The 5 dimensional physics around the UV region 
of the holographic coordinate should correspond to the UV physics of the dual field theory, and it is geometrically separated from the physics of the IR scales 
that are happening in the IR region of the holographic coordinate. This "locality" in energy scales seems to be an important peculiarity that holds in the strongly coupled, large $N_c$ gauge theory which has a holographic dual description \cite{Heemskerk:2009pn}. We assume that the large $N_c$ QCD in its low energy regime is one such theory, or more conservatively
we expect that a proper dual 5 dimensional theory may capture important physics of large $N_c$ QCD in its low energy, strongly coupled regime. Extending the results from large $N_c$ to $N_c=3$
is according to the usual spirit of considering the large $N_c$ approximation.

We will be interested in the high energy scattering with a low momentum transfer, $(-t)\equiv -q^2 \ll \Lambda_{QCD}^2$, or equivalently with a large impact parameter $b\gg \Lambda_{QCD}^{-2}$. 
Although the center of mass energy $\sqrt{s}$ is much larger than the QCD scale, the physics governing this scattering process is in the regime of "low energy", set by the momentum
transfer $\sqrt{-t}\ll \Lambda_{QCD}$, and it is non-perturbative, strongly coupled, warranting the application of ideas of AdS/CFT correspondence or holography.
There have been several works in this direction for both Pomeron and Reggeon exchanges \cite{Rho:1999jm,Janik:2000aj,Polchinski:2001tt,Brower:2006ea,Alday:2007hr,Barnes:2009ag,Nishio:2011xz,Giordano:2011sn,Giordano:2011ua,Basar:2012jb,Stoffers:2012zw,Watanabe:2012uc}. Here, we distinguish the amplitudes with Reggeon exchanges from those with Pomerons by
that Reggeons carry non-zero quantum numbers of the quark flavor symmetry whereas Pomerons have vacuum quantum numbers. 
A Reggeon then necessarily consists of at least a valence quark and an anti-quark pair in its wave function, something like a flavored meson. On the other hand, one may expect that
Pomerons are dominantly made of colorless, flavorless glueballs.
This naturally maps Reggeons to the holographic degrees of freedom describing flavored mesons (such as the 5 dimensional fields living on the "probe branes" representing
quark dynamics), and Pomerons to the bulk degrees of freedom corresponding to glueball dynamics.
The problem of high energy scattering in strong coupling would then be transformed to a scattering of the appropriate 5 dimensional degrees of freedoms in the 
holographic dual 5 dimensional space-time: for Reggeons, it will be an open string exchange amplitude, whereas for Pomerons, it would be a closed string amplitude.
Generally speaking, computing these amplitudes in a curved space-time such as the holographic dual space-time is hard and not well-known.
Various attempts with useful and meaningful approximations have been performed, which gave us fairly good, but still incomplete understanding on these amplitudes.
Since the description is supposed to capture full non-perturbative dynamics of QCD, the results should in principle satisfy all known consistency theorems such as the Froissart bound on the total cross section \cite{Froissart:1961ux}. 

We will follow the approximation introduced in Ref.\cite{Janik:2001sc} in our study of possible interplay between Reggeons and QED photons.
The idea of Ref.\cite{Janik:2001sc} is to approximate the full stringy amplitude with its semi-classical saddle point contribution. 
This approximation
can be well-justified in the framework of holography where the string world-sheet action is proportional to the large t'Hooft coupling factor $\sim\sqrt{\lambda}$ ($\lambda\equiv g_{YM}^2 N_c$) warranting
the use of semi-classical approximation. Moreover, it has been known that the high $s$, small $t$ regime of string amplitudes is generically governed by semi-classical world-sheet configurations, even without explicit strong coupling enhancement \cite{Gross:1987kza}. Based on this, one may expect that the semi-classical approximation captures the main ingredients of the scattering amplitudes in high $s$, small $t$ regime. Ref.\cite{Janik:2001sc} was able to describe the Regge behavior of the scattering amplitudes,
\be
{\cal T}\sim i\,s^{\alpha_0+\alpha' t}\,,
\ee
within this approximation, although there seem to be a few numerical mismatches with the full stringy computation in Ref.\cite{Basar:2012jb}. 

We will make a further approximation assuming that most of the string world-sheets in the semi-classical solutions reside at the IR bottom of the holographic coordinate with an effective string tension $T_{eff}={1\over 2\pi\alpha'}$. In the regime of interest $(-t)>0$ (space-like momentum transfer), the validity of this approximation is model-dependent as first discussed in 
Ref.\cite{Brower:2006ea} (see also Ref.\cite{Basar:2012jb} ). There is a general tendency of pushing the string world-sheets toward a more UV region when $(-t)>0$ whereas there is a competing gravitational attraction toward the IR region.
The former is proportional to $(-t)$ while the latter is independent of $(-t)$, so that for some moderate values of $(-t)$ the string world-sheets can indeed stay near the IR bottom in a model dependent way. Since we are going to focus on very small values of $(-t)$ ( in fact, $(-t)=0$ when we discuss the total cross section using the optical theorem), this "locality" to IR bottom
may well be justified. We should emphasize however that it would be certainly possible to include diffusion dynamics along the holographic coordinate beyond this approximation,
in a similar way to those done for the Pomeron amplitudes in Refs.\cite{Brower:2006ea,Nishio:2011xz,Stoffers:2012zw,Watanabe:2012uc}. 

\begin{figure}[t]
	\centering
	\includegraphics[width=8cm]{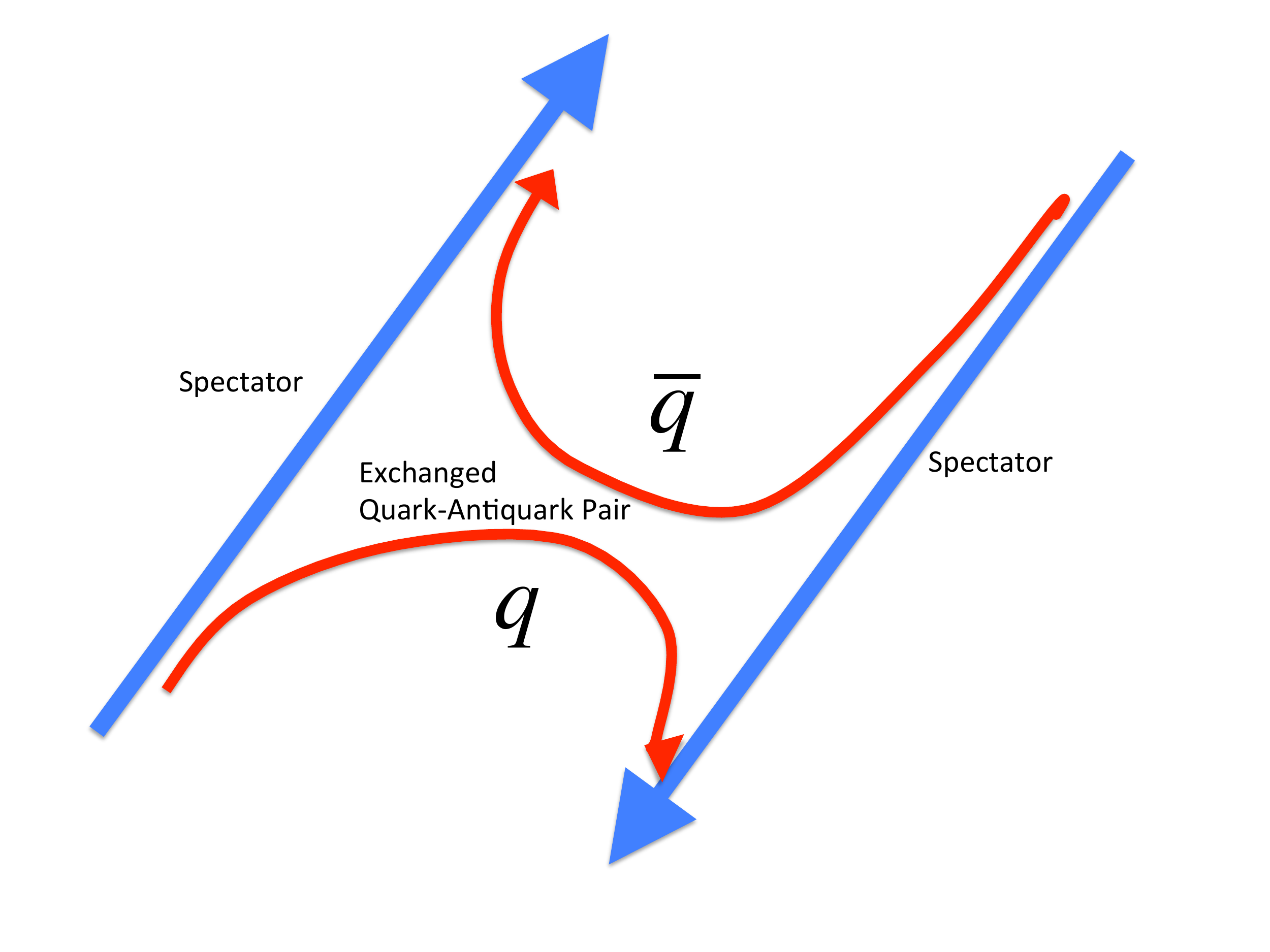}
		\caption{ Kinematics of a Reggeon exchange scattering in high energy eikonal limit. Spectators move straightly without deflection, while a quark-antiquark pair is exchanged. 
		A two dimensional open string world-sheet with the boundary set by the trajectories of exchanged quark-antiquark pair will describe the amplitude.  \label{fig00}}
\end{figure}
The kinematics of a Reggeon exchange amplitude is depicted in Figure \ref{fig00}.
Two projectiles carrying valence quarks or antiquarks exchange a pair of quark-antiquark pair during the collision process. In the high energy limit, 
the spectators other than the exchanged quark-antiquark pair travel straightly without much modification in their trajectories, and one can simply assume that their trajectories
are not modified at all in the eikonal approximation. 
The QCD in the strongly coupled, confined regime would results in configurations of QCD flux tubes (strings) joining the exchanged quark-antiquark pair, so that the space-time picture of the string world-sheet would be a two dimensional surface with its boundaries being the world-lines of the exchanged quark-antiquarks.  
The full string amplitude of this "open string" exchange would be obtained by summing over all configurations
of string world-sheets with the boundaries given by the world-lines of the exchanged pair.
Note that these boundaries, or the world-lines of the exchanged quark-antiquark pair are dynamical, so that the path integral of string configurations includes the variation
of the boundaries too. 
The asymptotic trajectories in the infinite past and the future, coinciding with the eikonalized spectator trajectories, set the boundary condition for this variational problem.
Note that while a Pomeron amplitude, which is described as a dipole-dipole scattering, can be reduced to a connected expectation value of two Wilson loops in the eikonal limit \cite{Nachtmann:1991ua}, our Reggeon amplitude, which involves an exchange of dynamical quark-antiquark pair, can not simply be reduced to Wilson lines/loops.
The task at hand is precisely equal to computing an open string amplitude, except the somewhat unconventional incoming/outgoing states specified in the real space-time rather than in the momentum space. 
We would like to approximate this amplitude with a semi-classical contribution obtained by finding an extremum solution of the classical Nambu-Goto string action
with the specified asymptotic boundary condition.

As in Ref.\cite{Janik:2001sc} we compute the amplitude in the Euclidean space and analytically continue the result to the Minkowski space-time.
The rapidity difference $\chi=\log s$ in the Minkowski space becomes an angle $\theta=-i\chi$ on the longitudinal plane in the Euclidean space~\cite{Meggiolaro:2007wn} between the two asymptotic trajectories of the projectiles.
The two asymptotic trajectories have an impact parameter $b$ in the transverse plane.
After identifying the dominant semi-classical configuration, the amplitude ${\cal T}(s,b)$ in the impact parameter space then is given by
\be
{{\cal T}(s,b)\over 2is}\sim {1\over s}\,\exp\left[-S_{\rm saddle}(\theta\to-i\chi)\right]\,,\label{t1}
\ee 
where the $1\over s$ in front is the spin factor arising from the Berry phase of the exchanged Dirac spinors in high energy limit \cite{Janik:2001sc}, and $S_{\rm saddle}$ is the classical Nambu-Goto string action evaluated on the extremum solution. 
The $t$-space amplitude ${\cal T}(s,t)$ is related to ${\cal T}(s,b)$ by the Fourier transform,
\be
{\cal T}(s,t)=\int d^2 b \,e^{iq\cdot b} \,{\cal T}(s,b)\quad,\quad t\equiv -q^2\,.\label{t2}
\ee 
Note that with our definition of the amplitude $\cal T$ as above, the total cross section via the optical theorem is given by
\be
\sigma_{\rm tot} = {1\over s}\, {\rm Im}{\cal T}(s,t=0)={1\over s}\int d^2 b\,{\rm Im}{\cal T}(s,b)={2\over s}\int d^2b\,{\rm Re}\left(e^{-S_{\rm saddle}(\theta\to-i\chi)}\right)(b)\,,
\ee
which we will use later.

\begin{figure}[t]
	\centering
	\includegraphics[width=10cm]{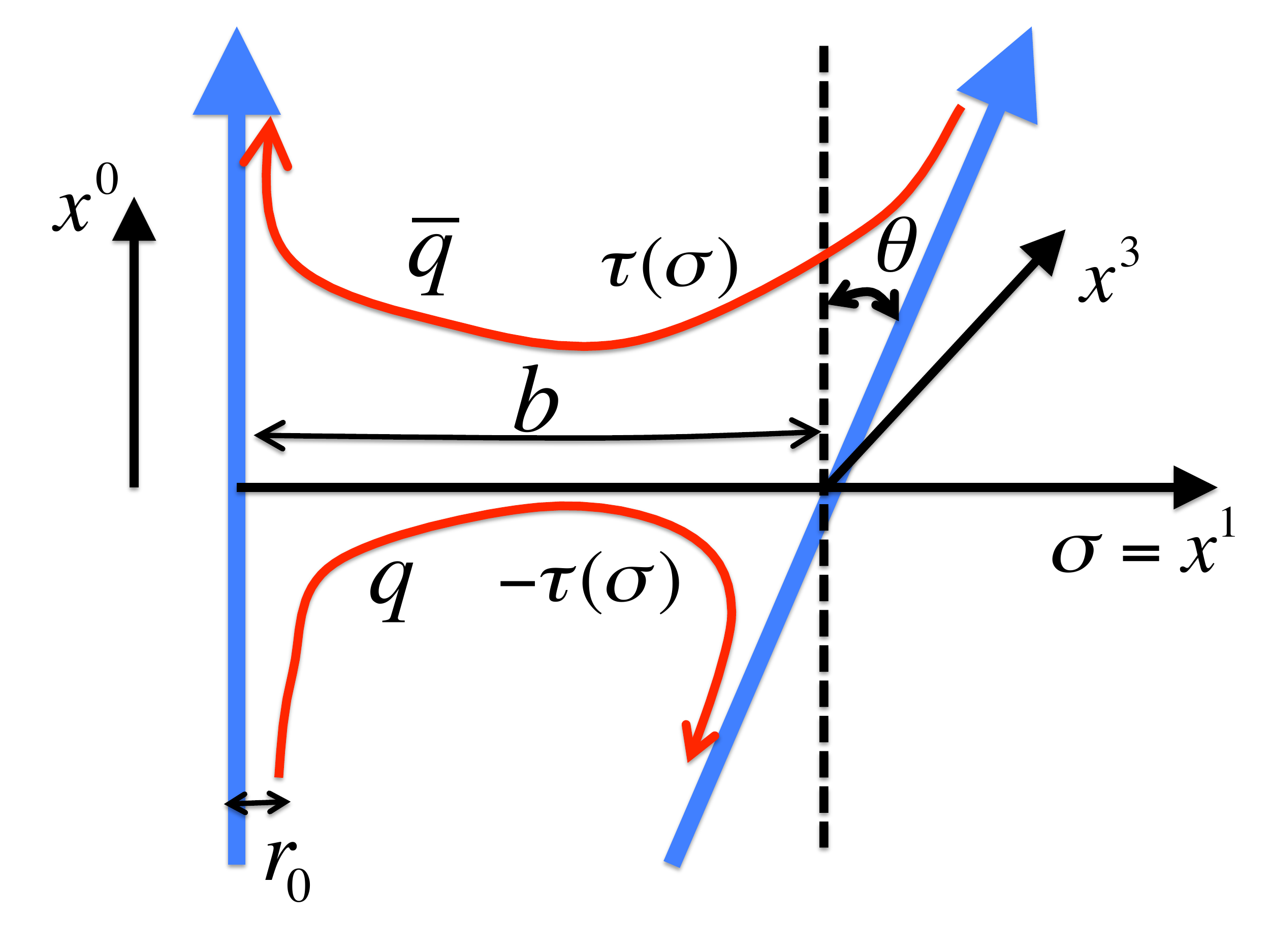}
		\caption{  Euclidean configuration of a high energy Reggeon scattering. The two spectator trajectories have a relative angle $\theta$ in the longitudinal plane $(x^0,x^3)$, and are separated by an impact parameter $b$ in the transverse plane $(x^1,x^2)$. The exchanged quark-antiquark pair world-lines reside on the helicoid spanned between these two straight lines of the spectators, and are parametrized by a single function $\tau(\sigma)$. \label{fig01}}
\end{figure}
Let the longitudinal plane be parametrized by $(x^0,x^3)$ and the world line of the first projectile be given by $(x^0,x^3)=(\tau,0)$ ($-\infty<\tau<\infty$), sitting at the origin of
the transverse plane spanned by $(x^1, x^2)$. The world line of the second projectile has a relative angle $\theta$ in the longitudinal plane, so that
$(x^0,x^3)=(\tau\cos\theta,\tau\sin\theta)$ ($-\infty<\tau<\infty$), and is separated by a distance $b$ in the transverse plane $(x^1,x^2)=(b,0)$.
We have to find the open string extremum whose asymptotic boundaries are specified by these two world lines (see Figure \ref{fig01}). Near each trajectories, the string world sheet
 tends to be parallel to the above straight lines, and since these two world lines have a relative angle $\theta$ with a distance $b$, the string world sheet has to twist
 itself on the longitudinal plane as it spans the transverse direction along $x^1$ of distance $b$ to connect the two world lines.
It is natural to expect that to a good approximation the string world sheet that minimizes its area lies close to the helicoidal surface bounded by the above two world lines,
\be
x^\mu(\sigma,\tau) : (x^0=\tau\cos(\theta(\sigma)),x^3=\tau\sin(\theta(\sigma)),x^1=\sigma, x^2=0)\quad,\quad \theta(\sigma)\equiv{\theta\over b}\sigma\,,
\ee
where the surface coordinates span $0<\sigma<b$ and $-\infty<\tau<\infty$.
This was pointed out in Ref.\cite{Janik:2001sc} and was successfully used to reproduce the leading Regge trajectory expected from the high energy limit of the open string Veneziano amplitude.
The open string world sheet on this helicoid surface corresponding to a Reggeon exchange will be specified by its two boundary curves on the helicoid surface joining the two straight
world lines, as these curves represent the world lines of the exchanged quark and antiquark pair. 
One can parametrize these curves in the helicoid surface coordinates $(\sigma,\tau)$ by a single function $\tau(\sigma)$ ($0<\sigma<b$), 
\be
(\sigma,\tau)=(\sigma,\pm\tau(\sigma))\,,
\ee
where the two curves differ simply by a sign of $\tau$ coordinate due to an obvious symmetry $\tau\to-\tau$. See Figure \ref{fig01} for details.
The open string world sheet therefore covers a part of helicoid surface specified by the coordinate range $0<\sigma<b$ and $-\tau(\sigma)<\tau<\tau(\sigma)$, with yet to be determined
function $\tau(\sigma)$ by extremizing the Nambu-Goto action.
The Nambu-Goto action is easily found to be
\be
S_{\rm Reggeon}= {1\over 2\pi\alpha'}\int_0^b d\sigma \int^{\tau(\sigma)}_{-\tau(\sigma)} d\tau\,
\sqrt{1+{\theta^2\over b^2}\tau^2}\,.
\ee
It is convenient to perform the analytic continuation $\theta\to-i\chi$ at this point to obtain
\be
S_{\rm Reggeon}= {1\over 2\pi\alpha'}\int_0^b d\sigma \int^{\tau(\sigma)}_{-\tau(\sigma)} d\tau\,
\sqrt{1-{\chi^2\over b^2}\tau^2}\,,
\ee
whereas we don't need to analytically continue the $\tau$ variable as it is just an integration variable\footnote{However, the solutions should be related by the analytic continuation, $\tau(\sigma)\to i \tau(\sigma)$.}. Extremizing the above gives an equation of motion
\be
1-{\chi^2\over b^2}\tau(\sigma)^2=0\,,
\ee
with the trivial solution $\tau(\sigma)={b\over\chi}$, whose on-shell action is
\be
S_{\rm saddle}= {1\over 2\pi\alpha'}\int_0^b d\sigma \int^{b\over\chi}_{-{b\over\chi}} d\tau\,
\sqrt{1-{\chi^2\over b^2}\tau^2}={b^2\over 4\alpha' \chi}\,.\label{puresol}
\ee
The resulting $t$-space amplitude is then given by
\be
{\cal T}(s,t)\sim 2i\int_0^\infty bdb\int_{0}^{2\pi}d\theta\,e^{i\sqrt{(-t)}\, b\cos\theta} e^{-{b^2\over 4\alpha'\chi}}\sim i (\log s)\, s^{\alpha' t}\,,
\ee
which shows the expected Regge behavior for Reggeon amplitudes \cite{Janik:2001sc}. In the next section, we will use this computational framework developed in Ref.\cite{Janik:2001sc} to 
study interesting interplay between Reggeons and QED photons in the regime where QED interactions can no longer be neglected due to a large $Z\sim 100$ enhancement for heavy-ions.

\section{Interplay of Reggeon and photon in pA collisions}

Within the framework of the previous section, we now include the QED interactions and establish how the QCD Reggeon
dynamics is affected by them.
There are two points that we have to emphasize. The first thing is that the effect we are discussing
is not a mere quantum mechanical interference between Reggeon and QED amplitudes when we square the total amplitude
to obtain cross-section. Rather, we are studying effects that modify Reggeon and QED amplitudes themselves due to an interplay
between the two. This clearly contrasts to the case of Pomerons, where the QED interactions do not affect Pomeron amplitudes {\it per se}
and the only effects one sees is the interference of mutually independent Pomeron and QED amplitudes in the cross-sections.
The difference between Pomerons and Reggeons in this aspect is due to the fact that Reggeons are in general charged under QED
interactions whereas Pomerons are neutral. This point will become clearer as we go on.
The second point we emphasize is about the magnitude of the QED interactions in the problem.
Previous works have ignored QED effects in computing Reggeon amplitudes relying on the smallness of $\alpha_{\rm EM}$
compared to the strong QCD interactions. This assumption however fails in our problem due to a large $Z\sim 100$ enhancement in $pA$ collisions, and one should take care of QED interactions properly in 
computing the full Reggeon and QED amplitudes.

\begin{figure}[t]
	\centering
	\includegraphics[width=10cm]{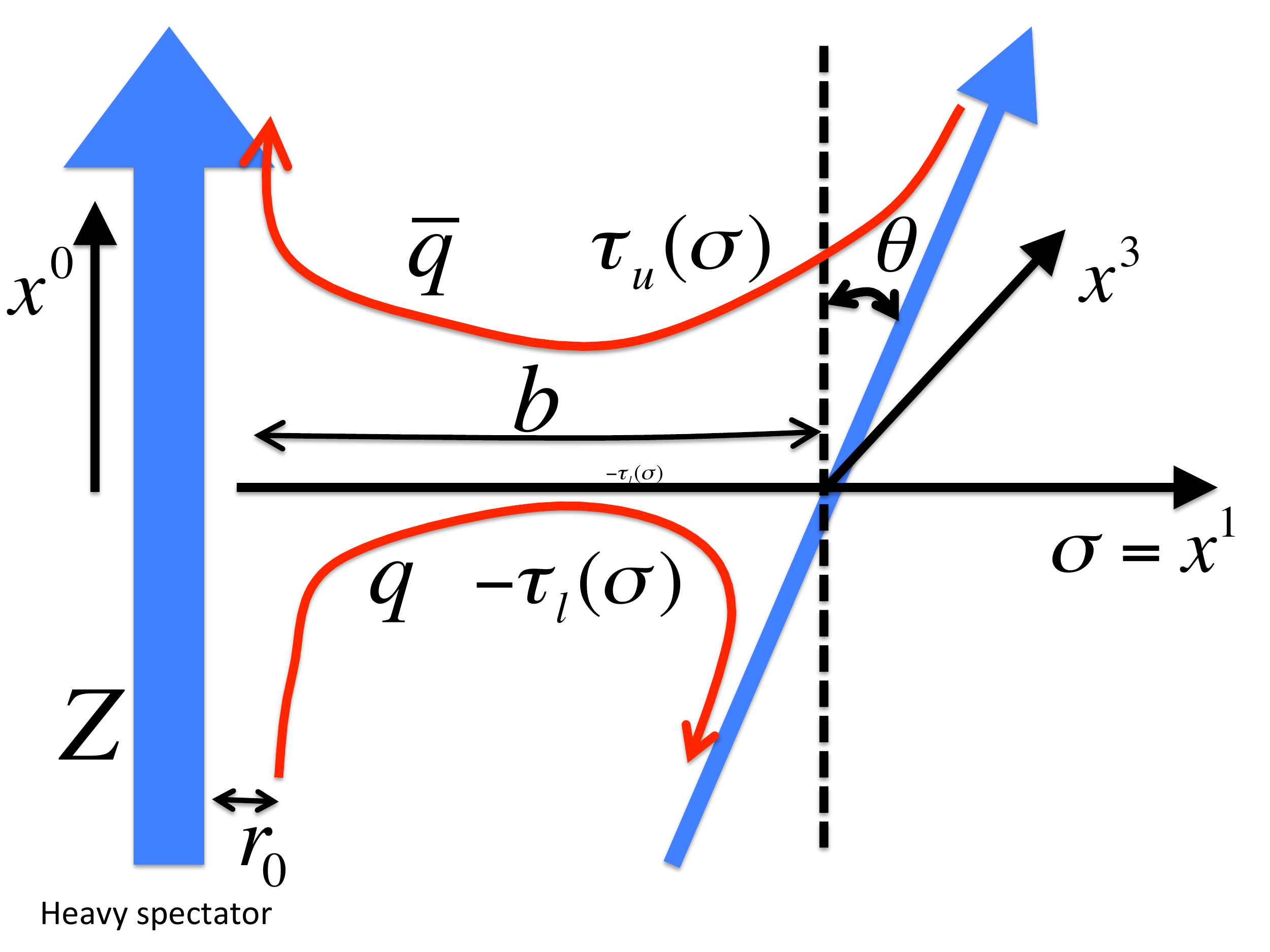}
		\caption{  Euclidean configuration of Reggeon exchange scattering between a heavy-ion with a charge $Z$ and a proton. Definitions are the same as in Figure \ref{fig01}, except that the exchanged quark-antiquark trajectories are parametrized by two independent functions $\tau_u(\sigma)$ and $\tau_d(\sigma)$. We include QED interactions between the charge $Z$ spectator and the rest of particles in leading large $Z$ approximation.\label{fig1}}
\end{figure}
We follow the same variational approach to the problem as described in the previous section. The trajectories of exchanged quark-antiquark pair
are assumed to be on the helicoid surface between the spectator trajectories. See Figure \ref{fig1} for the geometry and definitions of our problem. One difference is that in our case, the upper (quark) and lower (antiquark) trajectories are not necessarily symmetric with respect to time inversion $\tau\to-\tau$
when their QED charges are not equal in magnitude, so one has to introduce two separate functions $\tau_{u,d}(\sigma)$
to parameterize upper/lower trajectories. The range of worldsheet time $\tau$ for the string worldsheet 
bounded by exchanged quark-antiquark pair would then be
\bear
-\tau_d(\sigma)\le \tau\le \tau_u(\sigma)\quad.
\eear  
The non-perturbative QCD contribution (Reggeon contribution) to the action is as before
\be
S_{\rm Reggeon}= {1\over 2\pi\alpha'}\int_0^b d\sigma \int^{\tau_u(\sigma)}_{-\tau_d(\sigma)} d\tau\,
\sqrt{1+{\theta^2\over b^2}\tau^2}\quad.\label{Reggeon}
\ee

The QED part of the amplitude is simply the QED expectation value of the Wilson lines for charged particle trajectories
weighted by their charges. In leading large $Z$ approximation we are taking, the computation simplifies drastically
by that we only include mutual interactions between the charge $Z$ spectator and the rest particles.
In the approximation of neglecting dynamical quark-antiquark loop for photon propagator, the QED becomes a free 
theory of photons and the semi-classical treatment of Wilson line expectation values becomes exact.
In other words, the connected expectation value of multiple Wilson lines given by curves $C_i$,
\be
{\langle \prod_{i}W(C_i)\rangle \over \prod_i\langle W(C_i)\rangle}\quad,
\ee
is given by a product of connected expectation values of pairs of Wilson lines,
\be
\prod_{i\neq j} W_{ij}\quad, \quad W_{ij}\equiv{\langle W(C_i)W(C_j)\rangle \over \langle W(C_i)\rangle \langle W(C_j)\rangle}\quad.
\ee 
Also, $W_{ij}$ is computed simply by semi-classical expression
\be
W_{ij}=e^{iq_ie \int_{C_i}\, A^\mu_j dx_\mu}=e^{iq_je \int_{C_j}\, A^\mu_i dx_\mu}\quad,
\ee
where $A_i$ is the classical solution of Maxwell's equation sourced by the curve $C_i$ with charge $q_i$.

In leading $Z$ approximation we only have to compute $W_{ij}$ with $i$ being equal to the charge $Z$ spectator,
and it is sufficient to know the semi-classical $A_\mu$ sourced by the trajectory of charge $Z$ spectator which is
a straight line in (Euclidean) spacetime in high energy eikonal approximation. 
Working in the rest-frame of charge $Z$ spectator for convenience where it is located at spatial origin of the coordinate travelling straight along the (Euclidean) time, 
the solution is \footnote{We are working in Euclidean signature, and then will analytically continue to Minkowski spacetime
by $\theta\to -i\chi$, and $x^0\to it$, where $\chi$ is spacetime rapidity which should be roughly equal to the kinematic rapidity $\log\left(s\over m^2\right)$.}
\be
A_0=i{Ze\over 4\pi r}\quad,\quad r=\sqrt{\vec x\cdot\vec x}\quad,\quad \vec A=0\quad,\label{amu}
\ee
and the total QED amplitude in leading $Z$ approximation reads as
\be
\exp\left[ie\sum_i q_i\int_{C_i} A_\mu dx^\mu\right]\equiv \exp\left[-S_{\rm QED}\right]\quad,\label{qed}
\ee
where $i$ runs over rest of charged particles.
Without an interplay with the Reggeon contribution (\ref{Reggeon}), the QED amplitude in (\ref{qed}) would simply be a pure phase after analytic continuation to the Minkowski signature. As we will observe shortly, the interplay with Reggeons
can in general make QED amplitude to develop modulus change, that is, a nonzero real part in $S_{\rm QED}$.

It is easy to compute $S_{\rm QED}$ by integrating (\ref{amu}) over the trajectories of the rest particles.
See Figure \ref{fig1} for a pictorial explanation of the charged particle trajectories.
In the asymptotic past, one of the two incoming projectiles is a bound state of charge $Z$ spectator and a charge $q_d$ quark, 
and the latter will be exchanged to the other incoming projectile which comprises of a charge $-q_d$ antiquark and the spectator particle of charge $q$. We assume that inside the bound state, the charge Z spectator and the charge $q_d$ quark
are separated by a small distance $r_0$ which serves as a regularization of $S_{\rm QED}$ between the two. Our results are not sensitive to $r_0$. Phenomenologically it would be reasonable to take $r_0=1$ fm.
A small separation between the charge $-q_d$ antiquark and the charge $q$ spectator in the other projectile
is easily seen to be irrelevant and will be ignored. 
The same is true for the internal sizes of the charge $Z$ spectator and charge $q$ spectator, and we will treat them as point-like.
In the asymptotic future, the outgoing projectiles are a bound state of charge $Z$ spectator and a charge $q_u$ quark that has been exchanged, and a bound state of charge $q$ spectator and a charge $-q_u$ antiquark. 
We take the same assumption on the internal separation of constituents particles inside the bound states as in the asymptotic past.

We are only interested in the part of $S_{\rm QED}$ that depends on the trajectories of exchanged quark-antiquark pair 
given by two functions $\tau_{u,d}(\sigma)$, because we would like to perform a saddle point approximation
in integrating over those trajectories. We therefore won't consider the charge $q$ spectator in the following \footnote{Its contribution to the amplitude is simply a pure QED phase, some of which should be absorbed into
the asymptotic wavefunctions of incoming and outgoing projectiles.}.
The charge $q_d$ quark (or equivalently the charge $-q_d$ antiquark) trajectory given by $\tau_d(\sigma)$ (the lower trajectory in Figure \ref{fig1})
gives the contribution to $S_{\rm QED}$ as
\bear
S_{\rm QED}^{q_d}&=&
{q_d Z e^2\over 4\pi}\Bigg( {1\over r_0}\int_{-\infty}^{-\tau_d(0)} d\tau 
+\int_{-\tau_d(b)}^{-\infty}{d\tau\over\sqrt{\left(b+r_0\right)^2 +\tau^2\sin^2\theta}} \nonumber\\
&&-\int_0^b d\sigma
\,{\left(d\tau_d(\sigma)\over d\sigma\right)\cos\left(\theta\sigma\over b\right)-{\theta\over b}\tau_d(\sigma)\sin\left(\theta\sigma\over b\right)\over\sqrt{\left(\sigma+r_0\right)^2+\tau_d^2(\sigma)\sin^2\left(\theta\sigma\over b\right)}}\Bigg)\quad,\label{qd}
\eear
where the first line comes from the parts of asymptotic past (future) until (from) the Reggeon interaction region, and the second line
represents the contribution from the Reggeon exchange domain given by the interval $0 \le\sigma\le b$. See Figure \ref{fig1}.
The charge $q_u$ quark trajectory (the upper trajectory) gives a similar contribution
\bear
S_{\rm QED}^{q_u}&=&
{q_u Z e^2\over 4\pi}\Bigg( {1\over r_0}\int^{\infty}_{\tau_u(0)} d\tau 
+\int^{\tau_u(b)}_{\infty}{d\tau\over\sqrt{\left(b+r_0\right)^2 +\tau^2\sin^2\theta}} \nonumber\\
&&-\int_0^b d\sigma
\,{\left(d\tau_u(\sigma)\over d\sigma\right)\cos\left(\theta\sigma\over b\right)-{\theta\over b}\tau_u(\sigma)\sin\left(\theta\sigma\over b\right)\over\sqrt{\left(\sigma+r_0\right)^2+\tau_u^2(\sigma)\sin^2\left(\theta\sigma\over b\right)}}\Bigg)\quad.\label{qu}
\eear
The naive divergences in the first lines in (\ref{qd}) and (\ref{qu}) are simply either self-energy due to
QED interactions inside the asymptotic bound states, or the same divergences in usual QED pure phase of high energy eikonal scattering. The former is absorbed into the wavefunctions of incoming/outgoing states, while the latter can be regularized by an infrared (IR) cutoff which is also related to the definition of asymptotic wavefunctions. In any case, these divergences are independent of our variational $\tau_{u,d}(\sigma)$ and they are not of importance for our purposes.
We will simply regularize them by introducing an IR cutoff and replacing $\pm\infty\to\pm \Lambda_{\rm IR}$. 
Recall that upon analytic continuation to Minkowski signature, we will have to replace $\Lambda_{\rm IR}\to i \Lambda_{\rm IR}$.

The total Reggeon-QED action is a sum of (\ref{Reggeon}), (\ref{qd}), and (\ref{qu}), and the variational equations of motion
for $\tau_{u,d}(\sigma)$ can easily be obtained from them.
One can check that the boundary terms from the variations of second lines in (\ref{qd}) and (\ref{qu}) nicely
cancel with the variations of the first lines in (\ref{qd}) and (\ref{qu}), so there are no boundary conditions for $\tau_{u,d}(\sigma)$
at $\sigma=0,b$, in other words, their values and derivatives at the boundary $\sigma=0,b$ are unconstrained.
We will see in a moment that this is in fact consistent with the fact that the bulk equations of motion we get for
$\tau_{u,d}(\sigma)$ are algebraic, and we don't need and should not have any boundary conditions.

The equations of motion for $\tau_{u,d}(\sigma)$ do not mix with each other and we can treat them separately.
The equation of motion for $\tau_u(\sigma)$ reads 
\be
{1\over 2\pi\alpha'}\sqrt{1+{\theta^2\over b^2}\tau_u^2(\sigma)}
-{q_u Z e^2\over 4\pi}{\left(\cos\left(\theta\sigma\over b\right)\left(\sigma+r_0\right)+{\theta\over b}\sin\left(\theta\sigma\over b\right)\tau_u^2(\sigma)\right) \over\left(\left(\sigma+r_0\right)^2+\tau_u^2(\sigma)\sin^2\left(\theta\sigma\over b\right)\right)^{3\over 2}}=0\quad,
\ee
and the equation for $\tau_d(\sigma)$ is identical with $q_u\to q_d$. It is convenient to perform analytic continuation
$\theta\to -i\chi$ at this stage, and the equation of motion becomes upon defining $y(\sigma)\equiv \tau^2(\sigma)$
 (we omit subscript $u,d$ without much confusion)
\be
{1\over 2\pi\alpha'}\sqrt{1-{\chi^2\over b^2}y(\sigma)}
-{q_u Z e^2\over 4\pi}{\left(\cosh\left(\chi\sigma\over b\right)\left(\sigma+r_0\right)-{\chi\over b}\sinh\left(\chi\sigma\over b\right)y(\sigma)\right) \over\left(\left(\sigma+r_0\right)^2-\sinh^2\left(\chi\sigma\over b\right)y(\sigma)\right)^{3\over 2}}=0\quad.\label{master}
\ee
This is our master equation to solve for subsequent discussion.

\begin{figure}[t]
	\centering
	\includegraphics[width=10cm]{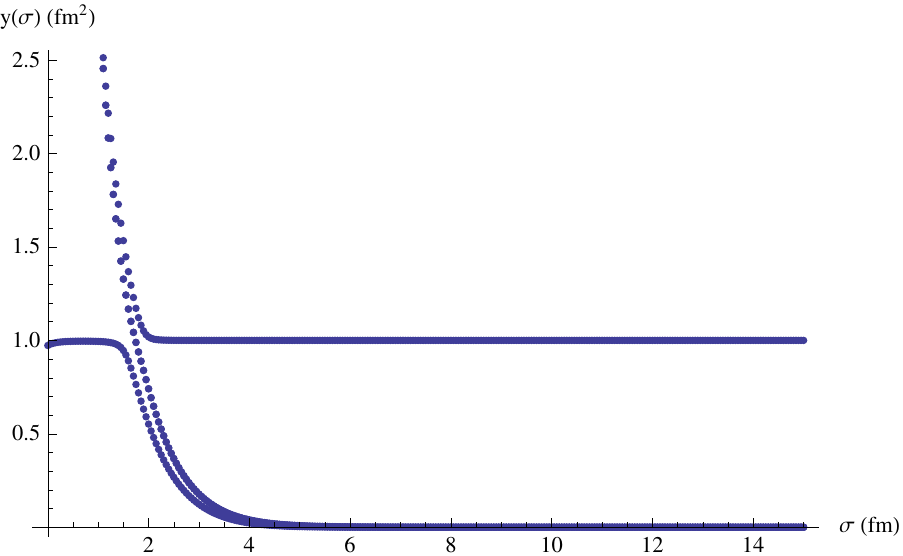}
		\caption{The numerical solution for (\ref{master}) with  $\alpha'=0.036\,{\rm fm}^{2}$, $r_0=1$ fm, ${e^2\over 4\pi}={1\over 137}$, $\chi=15$, $Z=100$, $q_u=1$, and ${b\over\sqrt{\alpha'}}=80$. One has to take the lower branch out of multiple solutions (saddle points). The transition happens when $\sigma_c\approx 1.73$ (see the text).  \label{fig3}}
\end{figure}
An example of numerical solutions is given in Figure \ref{fig3} for $\chi=15$, $Z=100$, $q_u=1$, and ${b\over\sqrt{\alpha'}}=80$, which shows
relevant generic features of the solutions for large $\chi$ and $b$. 
There exist two qualitatively different regions in the distance space $0\le\sigma\le b$ with a small transition region between them around $\sigma=\sigma_c$, where $\sigma_c$ is given by solving
\be
\left(\sigma_c+r_0\right)^2 = \sinh^2\left(\chi\sigma_c\over b\right)y(0)\quad,\label{cond}
\ee
with 
\be
y(0)=\tau(0)^2={b^2\over\chi^2}\left(1-\left(2\pi\alpha' q_u Z e^2\over 4\pi r_0^2\right)^2\right)\equiv {b^2\over\chi^2}\bar y(0)\quad,\label{tau0}
\ee
being the solution at $\sigma=0$.
As we see in Figure \ref{fig3}, $y(\sigma)$ is nearly constant in the region $0
\le\sigma\le\sigma_c$ with the value $y(0)$, and after that it quickly becomes exponentially small.
The condition (\ref{cond}) marks the point in $\sigma$ where the denominator in (\ref{master}) with $y(\sigma)\approx y(0)$
becomes zero, and one has a drastic change of behavior of the solution.
For large $\chi$ and $b$, one can find an approximate expression for the exponentially small behavior after $\sigma=\sigma_c$,
\be
y(\sigma)\approx -{\left({2\pi\alpha'q_u Z e^2\over 4\pi}\left(\sigma+r_0\right)\cosh\left(\chi\sigma\over b\right)\right)^{2\over 3}\over\sinh^2\left(\chi\sigma\over b\right)}\quad,\quad \sigma \gg\sigma_c\quad.\label{sol1}
\ee
Note that the solution is highly non-perturbative in QED coupling $\alpha_{\rm EM}={e^2\over 4\pi}$.
In fact, there is another branch of solution in the region $\sigma>\sigma_c$ (the upper branch in Figure \ref{fig3})
which is perturbative in QED interactions,
\be
y(\sigma)\approx {b^2\over\chi^2}+{\chi^4\over b^4}\left({2\pi\alpha'q_u Z e^2\over 4\pi}\right)^2\left({\cosh\left(\chi\sigma\over b\right)\left(\sigma+r_0\right)-{b\over\chi}\sinh\left(\chi\sigma\over b\right)\over\sinh^3\left(\chi\sigma\over b\right)}\right)^2\quad,\label{sol2}
\ee
and one has to choose the branch (\ref{sol1}) instead of (\ref{sol2}), as the former has a smaller value of real part of the
total action $S_{\rm total}=S_{\rm Reggeon}+S_{\rm QED}$.
For both branches (\ref{sol1}) and (\ref{sol2}), one can easily check that $S_{\rm QED}$ is purely imaginary for $\sigma\ge \sigma_c$, so that it doesn't play a role in finding the preferred saddle point \footnote{For the
branch (\ref{sol1}), $y(\sigma)<0$ so that $\tau(\sigma)$ is purely imaginary whereas the denominator in the second line of (\ref{qu}) is real, making its contribution to $S_{\rm QED}$ in (\ref{qu}) purely imaginary. On the other hand, $\tau(\sigma)$ in the branch
(\ref{sol2}) is real, but the denominator in (\ref{qu}) is purely imaginary, and $S_{\rm QED}$ is again purely imaginary. }. The real part of $S_{\rm Reggeon}$ which is simply
the area spanned by the string worldsheet then 
prefers smaller value of $y(\sigma)$, which is the branch (\ref{sol1}).

It is clear that the region $\sigma>\sigma_c$ with the preferred branch (\ref{sol1}) makes negligible contribution to the real part of $S_{\rm total}$ that we are interested in. It seems that the same is true for the small transition region around $\sigma\approx\sigma_c$, and
the dominant contribution comes from the interval $0\le\sigma\le \sigma_c$ with a nearly constant behavior of the solution
$y(\sigma)\approx y(0)$.
The contribution to the real part of $S_{\rm Reggeon}$ is therefore
\bear
&&{\rm Re}[S_{\rm Reggeon}]\nonumber\\
&=&{1\over 2\pi\alpha'}\int_0^{\sigma_c} d\sigma\int_0^{\tau(\sigma)} d\tau\,\sqrt{1-{\chi^2\over b^2}\tau^2}
\approx {\sigma_c\over 2\pi\alpha'}\int_0^{\tau(0)}d\tau\,\sqrt{1-{\chi^2\over b^2}\tau^2}\label{reReggeon}
\\
&=&{\sigma_c\over 4\pi \alpha'}{b\over\chi}\left(\left(2\pi\alpha' q_u Z e^2\over 4\pi r_0^2\right)\sqrt{1-\left(2\pi\alpha' q_u Z e^2\over 4\pi r_0^2\right)^2}+\sin^{-1}\left(\sqrt{1-\left(2\pi\alpha' q_u Z e^2\over 4\pi r_0^2\right)^2}\right)\right)\quad,\nonumber
\eear
using (\ref{tau0}) for $\tau(0)$. Numerically, $\left(2\pi\alpha' q_u Z e^2\over 4\pi r_0^2\right)\sim 0.132<1$ for reasonable values of parameters, such as $\alpha'=0.036\,\,{\rm fm}^2$, $Z=80$, $q_u=1$, $r_0=1\,\,{\rm fm}$, and ${e^2\over 4\pi}={1\over 137}$. It is interesting to note that for extremely large $Z$ such that $\left(2\pi\alpha' q_u Z e^2\over 4\pi r_0^2\right)>1$, $\tau(0)$ is imaginary and the whole $S_{\rm total}$ becomes purely imaginary.
We may interpret this as a complete QED domination over QCD Reggeon amplitudes, which however doesn't seem to happen in real
experiments.

The contribution from $S_{\rm QED}$ to the real part of $S_{\rm total}$ can also be obtained easily, after analytic continuation of $\Lambda_{\rm IR}\to i\Lambda_{\rm IR}$ and assuming ${d\tau(\sigma)\over d\sigma}\approx 0$ over $0\le \sigma\le \sigma_c$, 
\bear
&&{\rm Re}[S_{\rm QED}]=-{q_u Z e^2\over 4 \pi r_0}{b\over \chi}\sqrt{1-\left(2\pi\alpha' q_u Z e^2\over 4\pi r_0^2\right)^2}\nonumber\\
&+&{q_u Z e^2\over 4\pi}\sqrt{1-\left(2\pi\alpha' q_u Z e^2\over 4\pi r_0^2\right)^2}\int_0^{\sigma_c}d\sigma
\,{\sinh\left(\chi\sigma\over b\right)\over\sqrt{\left(\sigma+r_0\right)^2 - \sinh^2\left(\chi\sigma\over b\right)y(0)}}
\quad,\label{reqed}
\eear
where the first line comes from the boundary term at $\sigma=0$.

We are interested in the regime where $b$ is very large (equivalently a small Mandelstam variable $t$) in unit of Fermi.
In this case, the equation (\ref{cond}) for $\sigma_c$ is solved by
\be
\sigma_c=C{b\over\chi}\quad,
\ee
where $C$ is a ${\cal O}(1)$ numerical number determined by
\be
C^2=\left(1-\left(2\pi\alpha' q_u Z e^2\over 4\pi r_0^2\right)^2\right)\sinh^2(C)\quad.
\ee
For our previous parameters, $C\approx 0.23$. With this, the integral in the second line of (\ref{reqed}) can 
be performed to give 
\be
\int_0^{\sigma_c}d\sigma
\,{\sinh\left(\chi\sigma\over b\right)\over\sqrt{\left(\sigma+r_0\right)^2 - \sinh^2\left(\chi\sigma\over b\right)y(0)}}
=\int_0^C d\bar\sigma\,{\sinh\bar\sigma\over\sqrt{\bar\sigma^2-\sinh^2\bar\sigma \bar y(0)}}\quad,
\ee
in the ${b\over\chi}\to\infty$ limit, where a dimensionless number $\bar y(0)$ was defined previously in (\ref{tau0}).
The above integral is ${\cal O}(1)$ in $b\over\chi$, of a numerical value 2.74 with our previous parameters.
Since the contribution from the Reggeon action in (\ref{reReggeon}) is quadratic in $b\over\chi$, whereas the QED contribution
(\ref{reqed}) is at most linear in $b\over\chi$, the former dominates eventually for small $t$-channel exchanges.

In summary, the leading large $b$ asymptotics of the real part of $S_{\rm total}$ 
is
\be
{\rm Re}[S_{\rm total}]\approx {C\over 4\pi\alpha'}\left(\sqrt{\bar y(0)\left(1-\bar y(0)\right)}+\sin^{-1}\left(\sqrt{\bar y(0)}\right)\right){b^2\over\chi^2}+{\cal O}\left(b\right)\approx {1\over 4\pi\alpha^\prime(Z)}{b^2\over\chi^2}\quad,\label{mainres}
\ee
where $C$ is determined by
\be
C^2-\bar y(0)\sinh^2(C)=0\quad,
\ee
and $\bar y(0)$ depends on QED parameters as follows:
\be
\bar y(0)\equiv 1-\left(2\pi\alpha' q_u Z e^2\over 4\pi r_0^2\right)^2\quad.
\ee
This is our main result. One has to add the similar contribution from the lower part of the trajectory with a replacement $q_u\to q_d$.
From (\ref{mainres}) with the use of (\ref{t1}) and (\ref{t2}), we get the Reggeon amplitude 
\be
{\cal T}_{\rm Reggeon}^{Ap,A\bar p}(s,t)\sim i\, e^{\alpha^\prime(Z)t\,({\rm ln}s)^2}\,,
\ee
and the total cross section
\be
\sigma_{\rm Reggeon}^{Ap,A\bar p}\sim {(\log s)^2\over s}\,.
\ee

Let us now briefly discuss possible 1-loop contributions coming from summing over linearized fluctuations around the above saddle point solution in the path integral of string world-sheet.
The 1-loop determinant of massless fluctuations on the world-sheet gives the contribution to the action as
\be
S_{\rm 1-loop}={D_\perp\over 2}\log\det(\partial^2)\,,
\ee
where $D_\perp$ is the number of massless bosonic degrees of freedom on the string world-sheet\footnote{World-sheet Fermionic fields in general become massive and negligible
in the non-supersymmetric background of holographic QCD \cite{Kinar:1999xu}, indicating a transition from the critical superstring theory to an effective non-critical bosonic string theory \cite{Basar:2012jb}. This is a nice revival of the old string theory for QCD.}.
In the case of pure Reggeon exchange, the semi-classical string solution has a  rectangular shape whose length-sizes are $b$ and ${\pi\over 2}{ b\over \chi}$ (see the solution (\ref{puresol})).  
In the high energy limit, $\chi\to\infty$, the shape becomes highly elongated, $b\gg {\pi\over 2}{b\over\chi}$, and in this case the 1-loop action given above is dominated by the Casimir energy,
\be
{D_\perp\over 2}\log\det(\partial^2)\approx -{\pi D_\perp\over 24}{b\over \left({\pi\over 2}{b\over\chi}\right)}=-{D_\perp\over 12}\chi\,,
\ee
which enhances the scattering amplitude by a factor
\be
\exp\left(-S_{\rm 1-loop}\right)=e^{{ D_\perp\over 12}\chi}=s^{ D_\perp\over 12}\,.
\ee
As it is independent of $t$, it contributes to the intercept of the Regge trajectory which governs the large $s$ growth of the total cross section.
However, in our case of Reggeon interplay with QED photons, we see that our semi-classical string solution has the length-sizes of $\sqrt{y(0)}\approx {b\over\chi}$ and $\sigma_c\approx {b\over\chi}$ (since the solution collapses to an exponentially small size beyond $\sigma>\sigma_c$), and the shape remains symmetrical even in the high energy limit $\chi\to
\infty$. In this case, the 1-loop action is simply an order unity number, and we would not expect any large $\chi$ enhancement from it. 
Therefore, there wouldn't be a non-zero intercept from the 1-loop action, and one can simply take our main result (semi-classical contribution) as the dominant contribution
to the full amplitude in the large $\chi$ limit.
We emphasize that this qualitative change of leading high energy behavior is due to our non-trivial interplay with the QED photons which is non-perturbative in the QED coupling $\alpha_{\rm EM}$.

In summary, holographic Reggeon exchange scatterings in $pp$ or $p\bar p$ collisions, and 
holographic charged Reggeon exchange in $Ap$ or $A\bar p$ at large $Z$ compare as follows
\begin{eqnarray}
&&{\cal T}_{\rm Reggeon}^{pp,p\bar p}(s,t)\sim i\,s^{\alpha_0+\alpha^\prime t} \,,\nonumber\\
&&{\cal T}_{\rm Reggeon}^{Ap,A\bar p}(s,t)\sim i\, e^{\alpha^\prime(Z)t\,({\rm ln}s)^2}\,,
\end{eqnarray}
with $\alpha_0\approx D_\perp/12\approx 1/4$ and $\alpha^\prime(Z)$ as defined in~(\ref{mainres}). Empirically, the Reggeon intercept is $\alpha_0\approx 0.55$~\cite{Donnachie:1992ny}. 

\section{Conclusions}

We have found that the large charge of the nucleus in pA collisions modifies the Reggeon slope and its  intercept. The effect on the slope is large and leads to a shrinkage of the differential cross section of elastic and diffractive cross sections.  Although we have derived this effect in the context
of strong coupling holographic QCD, we expect it to be accessible also in perturbative QCD+QED  through a pertinent mixing
of the parton ladders with photon exchanges. A dedicated pA experiment at the LHC or RHIC using heavy nuclei with large Z will be able to test our results.

\vskip 1cm \centerline{\large \bf Acknowledgement} \vskip 0.5cm

We would like to thank Yuri Kovchegov, Edward Shuryak and Kirill Tuchin for discussions.
This work was supported by the U.S. Department of Energy under Contracts No.
DE-FG-88ER40388.

 \vfil

\end{document}